\documentclass[aps,prl,superscriptaddress,floatfix,twocolumn,a4paper,showpacs]{revtex4}
\usepackage{graphicx}
\usepackage{amsmath}
\usepackage{amssymb}
\usepackage{epsf,latexsym}
\usepackage{times}
\usepackage{epsfig}
\usepackage{citesort}

\def\be{\begin{equation}}
\def\ee{\end{equation}}
\def\bea{\begin{eqnarray}}
\def\eea{\end{eqnarray}}

\def\lsim{\raise0.3ex\hbox{$\;<$\kern-0.75em\raise-1.1ex\hbox{$\sim\;$}}}
\def\gsim{\raise0.3ex\hbox{$\;>$\kern-0.75em\raise-1.1ex\hbox{$\sim\;$}}}

\begin{document}
\title{$B\to K \pi$ Puzzle and New Sources of CP Violation in Supersymmetry}

\author{S. Khalil}
\affiliation{Centre for Theoretical Physics, The British
University in Egypt, El Sherouk City, Postal
No. 11837, P.O. Box 43, Egypt.} %
\affiliation{Department of Mathematics, Ain Shams University,
Faculty of Science, Cairo, 11566, Egypt.}
\author{A. Masiero}
\affiliation{Univ. of Padua, Phys. Dept."G. Galilei" and INFN, Section of
Padua, Via Marzolo 8, I-35131, Padua, Italy.}
\author{H. Murayama}
\affiliation{Institute for the Physics and Mathematics of the
Universe, University of Tokyo,
Kashiwa, Chiba 277-8568, Japan.} %
\affiliation{Department of Physics, University of California,
Berkeley, CA 94720, USA.} %

\date{\today }

\begin{abstract}
The difference between the CP asymmetries of the $B^0 \to K^+
\pi^-$ and $B^+ \to K^+ \pi^0$ decays has been recently confirmed
with an evidence larger than $5\sigma$'s. We discuss it as a
possible signal of new physics associated with new (large) CP
violation in the electroweak penguin contributions. We propose a
supersymmetry breaking scheme where such new sources of CP
violation  occur in the flavor non-universal trilinear scalar
couplings.
\end{abstract}

\maketitle

%
Establishing that the simple CKM pattern of the Standard Model
(SM), with its single CP violating phase, correctly accounts for
the vast and complex realm of hadronic flavor phenomena represents
a stunning milestone in our endeavor to understand fundamental
interactions. Yet, there exists a few sources of tension between
experimental data and SM predictions in $B$ physics~\footnote{With
the new $B_K$ lattice evaluation, the SM CKM may have difficulty
to correctly produce also the CP violation in the kaon system
\cite{buras}}, in particular, but not exclusively, related to CP
asymmetries in $b \to s$ transitions (see \cite{bona,soni08,hou08,
silvestrini} and for update \cite{lepton09}). Here we focus our
attention on the so-called "$B \to K\pi$ Puzzle". The recent Belle
and Babar Collaborations updates \cite{:2008zza} on the CP
asymmetry in the decays $B^0 \to K^+ \pi^-$ ( together with the
consistent CDF data) and $B^+ \to K^+ \pi^0$,

\bea%
{\cal A}_{CP}(B^0 \to K^+ \pi^-) &=& -9.7\pm 1.2 \%,\\
{\cal A}_{CP}(B^+ \to K^+ \pi^0) &=& +5.0\pm 2.5 \% ,%
\label{asymmetries}
\eea

confirm the existence of a non-vanishing difference between
the two CP asymmetries beyond $5 \sigma$ \cite{HFAG}:
\begin{eqnarray}
{\cal A}_{CP}(B^0 \to K^+ \pi^-) - {\cal A}_{CP}(B^+ \to K^+ \pi^0 )
\nonumber\\
=(14.7\pm 2.7)\% ~~~~ \label{difference}
\end{eqnarray}

In the SM with naive factorization these two asymmetries are
essentially equal. On the other hand, one expects that
the "improved"  BBNS QCD
factorization \cite{beneke} (QCDF) with appropriate $1/m_b$ corrections
( accounting for unknown final state interactions) \cite{neubert} may
(even largely) modify such naive expectation. However, even allowing
for a considerable freedom in choosing such corrections, QCDF fails to reproduce
the above experimental difference by several $\sigma$'s
\cite{khalil:2005qg,lunghi}. Also alternative approaches to QCDF, namely
perturbative QCD (PQCD)\cite{sanda} and Soft Collinear Effective Theory (SCET)
\cite{bauer}, are in trouble to satisfactorily reproduce the result in
Eq.\ref{difference}. Only fitting to the experimental data arbitrary contributions
corresponding to subleading terms in the power expansion, can one overcome the
$B \to K\pi$ puzzle; this latter approach, known as General Parametrization \cite{GP,
silvestrini}, gives up searching for a specific dynamics. Notice that whenever
dynamical assumptions are made, discrepancies between theory and experiment
in the $B \to K\pi$ CP asymmetries arise \cite{dynamics}.

Hence, even with all the caution needed in interpreting results related to
CP asymmetries in purely hadronic exclusive $B$ decays
\cite{martinelli,silvestrini},
one can certainly entertain the possibility that $B \to K \pi$ puzzle
hints at some  physics beyond the SM  with new sources of
 CP violation in addition to the CKM phase \cite{KPI,baek}.

As said above, it is true that the $B \to K\pi$ puzzle is not the only potential
hint for new physics that emerges from rare B decays. Some tension among
values of the parameter $sin2\beta$ which are extracted in different ways
 from the data has been persisting for some time now and, more recently,
a possible evidence for new physics in the CP violating $b \to \psi \phi$
decay has been pointed out \cite{bona}. However, the anomaly in the CP
asymmetries of the two isospin-related B decay channels in Eq.\ref{asymmetries}
looks peculiarly interesting for the following aspect.
The $B$ decay amplitudes into $K^+ \pi^-$ and $K^+ \pi^0$ differ only by the
subleading terms given by the color-suppressed tree contribution ($C$) and
the electroweak penguins ($P_{EW}$). While a resolution of the $B \to K\pi$
puzzle through an enhancement of the $C$ amplitude is unviable \cite{baek},
prospects look more appealing if one tries to invoke a large CP violation
from $P_{EW}$ \cite{baek,peskin}.

On the other hand, $P_{EW}$ is
essentially real within
the SM and has a strong phase very close to the color-allowed
tree amplitude ($T$) \cite{rosner}. Hence, making use of  $P_{EW}$ for the
resolution of the $B \to K\pi$ puzzle
 entails the presence of new physics beyond the SM
with the presence of new sources of CP violation leaking into the electroweak
penguins. Indeed, we emphasize that what is actually crucial to overcome the
"$B \to K\pi$ puzzle" is that $P_{EW}$ exhibits a large CP violation, but
otherwise the new physics electroweak penguins need not be strongly enhanced
with respect to the SM ones.
This latter observation plays a major role when one tries to account for
the experimental result in Eq. (\ref{difference}) invoking new physics
 while respecting the vast set
of data in rare $B$ physics which represent a stunning confirmation of the
SM flavor paradigm encoded in the CKM matrix.

Here we address the above issue in the context of supersymmetric (SUSY)
extensions of the SM. If SUSY appears in a context where flavor physics is
still fully accounted for by the CKM pattern ( the so-called "minimal flavor
violation"), then the above CP puzzle remains untouched. It was pointed out
that prospects change when one moves to non-minimal flavor
SUSY models where the electroweak penguins can be enhanced and new phases
may be obtained \cite{khalil:2005qg}. It is then compelling to provide
an explicit model where the CP puzzle in $B \to K \pi$ decays is overcome
and to study its implications
for the flavor changing neutral current (FCNC) phenomenology. This is the aim
of the present paper.

We individuate the source of flavor non-minimality in the trilinear scalar
terms of the soft SUSY breaking sector, i.e. in the non-universal $A$-terms.
Non-universality in the
soft breaking terms is a common feature in most superstring
inspired SUSY models \cite{Brignole:1993dj}. Indeed, asking
for flavor universality in the SUSY breaking sector of supergravities
implies strong constraints on their  minimal Kahler potential. Relaxing such
constraints, as in several string and D-brane derived models
\cite{Khalil:2000ci}, leads to non-degenerate trilinear
couplings. It is also worth reminding that non-universal soft SUSY breaking
represents an important ingredient, together with new large SUSY CP phases, to
produce observable effects in the low-energy CP violating phenomena
without exceeding the tough constraint of the experimental electric dipole moments
limits \cite{Khalil:2000ci,Abel:2001vy}.

%
The $B\to K\pi$ decays are driven by the $b \to s $ transition.
The effective Hamiltonian of this transition is given by
\begin{eqnarray}
H^{\Delta B=1}_{\rm eff}&\!=\!&\frac{G_F}{\sqrt{2}} \sum_{p=u,c}
\lambda_p \Big(C_1 Q_1^p + C_2 Q_2^p + \sum_{i=3}^{10}C_i Q_i
\nonumber\\
\!\!&\!+\!&\! C_{7\gamma} Q_{7\gamma} + C_{8g}
Q_{8g}\Big)+\left\{Q_i\to \tilde{Q}_i\, ,\, C_i\to
\tilde{C}_i\right\}~~~~ \label{Heff}
\end{eqnarray}
where $\lambda_p= V_{pb} V^{\star}_{p s}$, with $V_{pb}$ the
unitary CKM matrix elements satisfying the unitarity triangle
relation $\lambda_t+\lambda_u+\lambda_c=0$, and $C_i\equiv
C_i(\mu_b)$ are the Wilson coefficients at low energy scale
$\mu_b\simeq {\cal O}(m_b)$. The basis $Q_i\equiv Q_i(\mu_b)$ of
the relevant local operators renormalized at the same scale
$\mu_b$ can be found in Ref.\cite{Buchalla:1995vs}.
$Q^p_{1,2}$ refer to the current-current operators, $Q_{3-6}$ to
the QCD penguin operators, and $Q_{7-10}$ to the electroweak
penguin operators, while $Q_{7\gamma}$ and $ Q_{8g}$ are the
magnetic and the chromo-magnetic dipole operators, respectively. In
addition, the operators $\tilde{Q}_{i}\equiv \tilde{Q}_{i}(\mu_b)$ are
obtained from $Q_{i}$ by the chirality exchange $(\bar{q}_1
q_2)_{V\pm A} \to (\bar{q}_1 q_2)_{V\mp A}$. Notice that in the SM
the coefficients $\tilde{C}_{i}$ identically vanish due to the V-A
structure of charged weak currents, while in the Minimal
Supersymmetric SM (MSSM) they can receive
contributions from both chargino and gluino exchanges.

In our analysis, we adopt the QCDF scheme \cite{beneke} in
evaluating the hadronic matrix elements for exclusive hadronic
final states. We assume that the QCD factorization free parameters
$\rho_{A,H}$ and $\phi_{A,H}$ are of order one. In this case, the
SM contributions to the amplitudes of $B^0 \to K^+ \pi^-$ and $B^+
\to K^+ \pi^0$ can be parameterized as \cite{neubert}
\begin{eqnarray}
A(B^0\to K^+ \pi^-)&=&\lambda_u T e^{i\delta_T} + \lambda_c \left( P e^{i\delta_P}+ P_{EW}^C e^{i\delta_{EW}^C}\right), \nonumber\\
\sqrt{2} A(B^+\to K^+ \pi^0 ) &=& \lambda_u \left( T
e^{i\delta_T}+ C e^{i\delta_C}\right) \nonumber\\
&+& \lambda_c \left(P e^{i\delta_P}+ + P_{EW}
e^{i\delta_{EW}}\right),~~~ %
\label{parametrization1}
\end{eqnarray}
where the real parameters: $T,C,P, P_{EW}$, and $P_{EW}^C$
represent  color allowed tree,  color suppressed tree, QCD
penguin, electroweak penguin, and color suppressed electroweak penguin
diagrams, respectively. The parameters $\delta_{P,T,C,EW,EW^C}$
denote the CP conserving (strong) phases of the corresponding amplitudes.
In the above expressions, we
neglected the small annihilation contribution. Note that in the
SM the only source of CP violation for both the B decays under consideration
is represented by the phase of the CKM parameter $\lambda_u$, ie. the penguin
contributions are essentially real. As stated above,
one finds that the SM contributions to the
CP asymmetries of $B^0\to K^+ \pi^-$ and
$B^+\to K^+ \pi^0$  are very close and with the same sign. Even
 assuming that the QCD factorization
parameter $\rho \gg 1$,  the two asymmetries still remain
of the same sign \cite{khalil:2005qg}, at variance with the present
experimental result.

%
We now focus our attention on SUSY models. The relevance of
$b - s$ transition processes as powerful probes for the presence
of low-energy SUSY has been long studied for the most general
class of MSSM \cite{ciuchini:bs,silvestrini}. Here
we stick to the interesting class of MSSM realizations where the
source of flavor non-universality resides entirely in the
non-universal $A$-terms.
The relevant Wilson
coefficients due to the gluino exchange are
$C_{7\gamma}^{\tilde{g}}$ and $C_{8g}^{\tilde{g}}$, while for the
chargino contribution the following Wilson coefficients are
important: $C_{7}^{\chi}, C_{9}^{\chi},
C_{7\gamma}^{\tilde{\chi}}$,and $C_{8g}^{\chi}$. Note that gluino
contributions to $C_7$ and $C_9$ are very small, particularly in
this class of model with non-universal $A$-terms and universal
squark masses. The complete expressions for these Wilson
coefficients can be found in terms of mass insertion approximation
in Ref.\cite{khalil:2005qg, gabrielli} and in terms of mass eigenstate in
Ref.\cite{Kruger:2000ff}. In our numerical analysis, we use the
complete one-loop computation in the mass eigenstate basis.

While for $T$, the SM contribution,
$T=T_{SM}$, dominates, sizeable SUSY contributions arise in the
penguin sector, $P = P_{SM} + P_{SUSY}$.
Notice that the strong CP violating phases associated
with $T_{SUSY}$ and $P_{SUSY}$ are in general different from the
SM ones.

As said above, we are going to make use of
 $P^{EW}_{SUSY}$ to account for the isospin breaking
difference in the CP asymmetries of Eq. (\ref{difference}). It is then
relevant to observe that
the electroweak penguins of $B\to K \pi$ are
insensitive to the values of $C_{7\gamma}$ and $C_{8g}$, which
give the dominant contributions to the branching ratio of $b \to s
\gamma$. Therefore, the $b\to s \gamma$ constraint
does not pose an immediate threat to our proposal.


In general, the SM and SUSY $B\to K\pi$ decay amplitudes can be parametrized
as follows: %
\bea %
A^{\mathrm{SM}} &=& \vert A^{\mathrm{SM}} \vert ~ e^{i(\theta^{\mathrm{SM}}+\delta^{\mathrm{SM}})},~\nonumber\\
\bar{A}^{\mathrm{SM}} &=& \vert A^{\mathrm{SM}} \vert
~ e^{i(-\theta^{\mathrm{SM}}+\delta^{\mathrm{SM}})},~~%
\eea%
with similar expression for $A^{\mathrm{SUSY}}$. Here,
$\delta^{\mathrm{SM(SUSY)}}$ is the SM (SUSY) CP conserving phase
and $\theta^{\mathrm{SM(SUSY)}}$ is the SM (SUSY) CP violating
phase.

The CP asymmetry can be written as
\be %
A^{CP} = \frac{2 R \sin(\delta_{SM}-\delta_{SUSY})
\sin(\theta_{SM} -\theta_{SUSY})}{1+R^2 + 2 R
\cos(\delta_{SM}-\delta_{SUSY}) \cos(\theta_{SM}
-\theta_{SUSY}) },%
\ee%
where $R$ is defined as $R=\vert A^{SUSY}/A^{SM} \vert$.

Let us turn to a specific model where to compute the relevant quantities
entering the
above expression for $A^{CP}$. We consider a SUSY breaking mechanism giving rise
to flavor non-universal
trilinear couplings.  We parameterize the trilinear matrices
$Y^{A}_{u,d}$ with the so-called "factorizable" $A$-terms,
$(Y A)_{ij} = A_{ij}Y_{ij}$,
where $A_{ij}$ is given by%
\be %
A^u =A^d =  \tilde{m}_0 \left(\begin{array}{ccc}
    x & y & z \\
    x & y & z \\
    x & y & z \end{array}
\right),%
\ee %
with the entries $x, y$ and $z$  complex and of order one, while
 $Y_{ij}$ are the Yukawa couplings.        In
general $A^d$ and $A^u$ have different structures. Here we assume
for simplicity that $A^d = A^u$. Again just for simplicity,
we consider universality both in the soft
scalar mass $\tilde{m}_0$ and  gaugino mass $M_{1/2}$ sectors.

In the super-CKM basis $Y^A$ reads $Y^A= Y_{diag}.
(U.A.V)$, where $U$ and $V$ are the left and right rotational
matrices that diagonalize the quark mass matrix.
The off-diagonal term in the $LR$ squark mass matrix is proportional to
the corresponding quark mass

\be %
(\delta^q_{LR})_{ij} = \frac{m^q_i}{\tilde{m}_q^2}
\left(U.A^q.V\right)_{ij},%
\ee %
where $\tilde{m}_q$ denotes the average squark mass.
Notice that the above choice of the "factorizable" $A$-terms
implies that the mass insertion $(\delta^u_{LR})_{11}
\simeq m_u/\tilde{m}_0 \sim {\cal O}(10^{-6})$, which is
consistent with the stringent neutron EDM constraint \cite{Abel:2001vy},
even in the presence  of large SUSY CP violating phases.

 The possibility of exploiting complex non–universal
$A$-terms had been advocated a few years ago in
connection with a similar issue of direct CP violation, but
in the context of the Kaon system, to account for the size of
$\varepsilon^\prime/\varepsilon$ \cite{Masiero:1999ub}.
Notice that in the present case involving transitions from
the third to the second generation, the size of
the flavor changing mass insertions can be quite conspicuous. Indeed,
$(\delta^u_{LR})_{32}$, which is relevant for the CP asymmetry of
$B \to K \pi$ mediated by chargino exchange, can be as high as
 $(\delta^u_{LR})_{32} \simeq
m_t/\tilde{m}_0 \sim {\cal O}(0.1)$. Clearly, the corresponding
$(\delta^d_{LR})_{32}$ in the down-sector is suppressed by the
smallness of $m_b$ compared to $m_t$ avoiding possible problems
with the $b\to s \gamma$ constraint.

\begin{figure}
\begin{center}
\epsfig{file=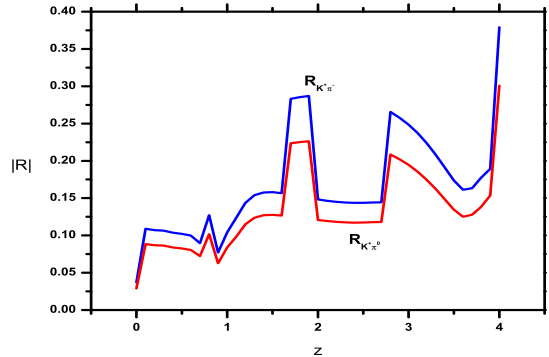, width=8cm,height=5.5cm,angle=0}
\end{center}
\vspace{-0.2cm} \caption{The ratios $R_{\pi^+ K^-}$ and $R_{\pi^0
K^-}$ as function of the trilinear parameter $z$ for $x=1$, $y=2$
and $\phi_i= {\cal O}(1)$.} \label{fig:R24}
\end{figure}
We now come to a quantitative evaluation
of the SUSY contributions to the CP asymmetries in our model.
In addition to $\tan \beta$, the free parameters are: $m_0,
m_{1/2}, A_0, \vert x \vert , \vert y \vert , \vert z \vert ,
\phi_1, \phi_2, \phi_3$, where $\phi_i$ are the associated phases
to the trilinear parameters $x,y,z$. For simplicity, we assume
$A_0 = m_0$ and set $m_0=300$ GeV, $m_{1/2} = 500$ GeV and $\tan
\beta=10$. We vary the other six parameters, taking into
account all the relevant constraints imposed on SUSY models, in
particular the EDM constraints and the observed limits on the
branching ratio of $b\to s \gamma$. These two constraints are the
most important ones to significantly affect our parameter space.

In order to account for the experimental results we ask
for the SUSY contributions to be such to give rise to the
following two features at variance with respect to what occurs in the SM:
i) $R_{\pi^+ K^-}$ should turn out to be larger than
$R_{\pi^0 K^-}$, in order to obtain  $ \vert A_{CP}(\pi^+
K^-)\vert > \vert A_{CP}(\pi^0 K^-)\vert$ and ii) the relative
sign between the CP violating and strong
phases associated to $B^0\to \pi^+ K^-$ should be negative, while
it should be positive in the  $B^-\to \pi^0 K^-$ case.

In Fig. 1 we display the ratios $R_{\pi^+ K^-}$ and $R_{\pi^0
K^-}$ as function of the trilinear parameter $z$ for $x=1$, $y=2$
and $\phi_i= {\cal O}(1)$ (other parameters are fixed as stated
above). As can be seen from this figure, the values of $R_{\pi^+
K^-}$  is typically larger than the values of the
ratio $R_{\pi^0 K^-}$.

We have checked that for the above choice of the parameter
 also the second above requirement, namely
the fact that the the CP violating phases of
the amplitudes
$B^0 \to \pi^+ K^-$ and $B^- \to \pi^0 K^-$ have opposite sign, is indeed
fulfilled.

\begin{figure}
\begin{center}
\epsfig{file=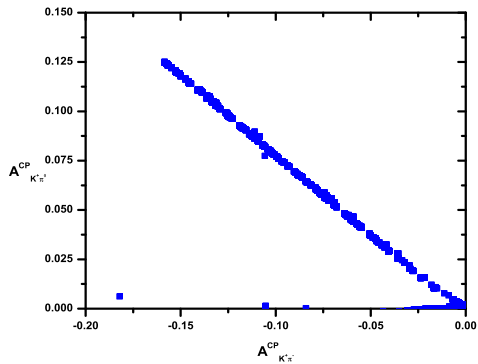, width=7cm,height=5.5cm,angle=0}
\end{center}
\vspace{-1cm} \caption{The correlation between the CP asymmetries
$A^{CP}(\pi^+ K^-)$ and $A^{CP}(\pi^0 K^-)$ as function of the
trilinear parameter $z$.} \label{fig:corr}
\end{figure}

The correlation between the two CP asymmetries $A^{CP}(\pi^+ K^-)$
and $A^{CP}(\pi^0 K^-)$ are displayed in Fig. 2.

From this figure,
it becomes clear that the experimental results reported in Eq. (\ref{asymmetries})
can be easily accommodated in this class of SUSY models.

In conclusion, we have shown that the presence of new sources of
CP violation in the  flavor non-universal trilinear scalar terms
can originate a ( large) phase in the SUSY penguins, in particular
in the electroweak ones, leading to a resolution of the $B \to
K\pi$ puzzle. Interestingly enough, these sources of CP violation
which are linked to the breaking of SUSY could be at the origin of
possible deviations from the SM in the two manifestations of
direct CP violation in the $K$ and $B$ systems,
$\varepsilon^\prime/\varepsilon$ and the $B \to K\pi$ CP violating
asymmetries, respectively. The experimental search for other B
decay channels ( for instance, $B^+ \to J/\psi K^+$ \cite{hou08})
where the presence of the enhanced CP violation in electroweak
penguins can show up is of utmost relevance.

$\it{Acknowledgements}$  We thank the Yukawa institute (Kyoto) and
the YKIS research project for kind hospitality and support while
part of this project was accomplished. S.K. would like to thank
the INFN Padua Section and the Department of Physics of the Padua
University for their kind hospitality while finishing this work.
S.K. work is partially supported by ICTP project 30 and the
Egyptian Academy of Science and Technology. A.M. acknowledges
partial support from RTN European Contracts MRTN-CT-2006-035863
(UniverseNet), MRTN-CT-2004-503369 ( QFU) and MRTN-CT-2006-035505
(HepTools), from the AstroParticle Special Projects of
 MIUR (PRIN 2006023491-001) and INFN (FA51), and the CARIPARO Excellence Project LHCosmology.

\end{document}